\documentclass[reprint,twocolumn,nofootinbib]{revtex4-1}

\usepackage[english]{babel}
\usepackage{graphicx} 
\usepackage{float}
\usepackage{amsmath}
\usepackage{amssymb}
\usepackage{algcompatible,algpseudocode}
\usepackage{newfloat}

\DeclareFloatingEnvironment[
    fileext=loa,
    listname=List of Algorithms,
    name=Algorithm,
    placement=tbhp,
]{algorithm}
\begin{document}

\title{Nonlinear Volterra Model of a Loudspeaker Behavior\\Based on Laser Doppler Vibrometry}
\author{Alessandro Loriga}
\affiliation{Intranet Standard GmbH, Ottostrasse 3, 80333 Munich, Germany}
\author{Parvin Moyassari}
\affiliation{Intranet Standard GmbH, Ottostrasse 3, 80333 Munich, Germany}
\author{Daniele Bernardini}
\affiliation{Intranet Standard GmbH, Ottostrasse 3, 80333 Munich, Germany}
\author{Gregorio Landi}
\affiliation{Dipartimento di Fisica, Universit\`a di Firenze, Italy, and INFN, Sezione di Firenze, Largo Enrico Fermi 2, Firenze 50125, Italy}
\author{Elisabeth Dumont}
\affiliation{Institute of Applied Mathematics and Physics, Zurich University of Applied Sciences, Technikumstrasse 9, 8401 Winterthur, Switzerland}
\author{Francesca Venturini}
\affiliation{Institute of Applied Mathematics and Physics, Zurich University of Applied Sciences, Technikumstrasse 9, 8401 Winterthur, Switzerland}

\begin{abstract}
We demonstrate the capabilities of nonlinear Volterra models to simulate the behavior of an audio system and compare them to linear filters. In this paper a nonlinear model of an audio system based on Volterra series is presented and Normalized Least Mean Square algorithm is used to determine the Volterra series to third order. Training data for the models were collected measuring a physical speaker using a laser interferometer. We explore several training signals and filter's parameters. Results indicate a decrease in Mean Squared Error compared to the linear model with a dependency on the particular test signal, the order and the parameters of the model.

\end{abstract}

\maketitle

\section*{INTRODUCTION}
It has been shown that physical loudspeaker are causal system \cite{echocancellation}. A physical loudspeaker has inherent nonlinearities generating
harmonic and intermodulation distortions in the output signal. The usual approach to correct audio signals is applying a linear filter due to computational advantages and simplicity.
Unfortunately, this approach leads to a magnitude amplification of the harmonics introduced by the nonlinear components. This means that the first order term of the system is well corrected while the error on the nonlinear terms becomes larger \cite{assessmentlinear}.\\
To overcome this problem one can use Volterra series to model the nonlinear behavior of the audio system \cite{echocancellation, klippelnonlinear, adaptforecho, lmsvolterra}. In this paper we use Volterra series up to third order \cite{nonlinearvolterra, nonlinearidentification, measuringvolterra, practicalmodeling}. The series kernel will be estimated through a learning algorithm. We will then compare the result with the linear filter approach.
The Normalized Least Mean Square (NLMS) algorithm will be used to estimate the first three orders of the Volterra series, using input and output signals measured with Polytec's Laser Doppler Vibrometer. The choice of a laser interferometer, as compared to standard electro acoustic measurement with a microphone, offers the possibility to characterize the membrane dynamics directly with high precision measurement of the acoustic velocity. This instrument performs an analysis under simpler conditions, since free field or anechoic chamber have not been available. In this paper we do not consider the effects resulting from breakout modes or planar waves on the membrane. To account for these effects one could use more than one interferometer pointed at the membrane and calibrate the response of the chosen output signal. Although this would be needed for an application, we did not deem it necessary to validate the methodology presented here.
The results are presented as Mean Square Error (MSE) between the measured output signal and those obtained with the first, second and third order model. MSE has been chosen due to its ability to represent the general optimization of the system, so it is easy to evaluate the method. Different signals are tested in order to study the error in different situations. We show that the error decreases when higher orders of the model are taken into account. Moreover a comparison is made between different signals used during training phase.
\section{The Volterra Model}
A discrete causal time-invariant system with memory can be represented by the Volterra series as, \cite{volterrabook}:
\begin{equation}
y(n) = h_0 + \sum_{p=1}^\infty \sum_{\tau_1=a}^b ... \sum_{\tau_p=a}^b \! h_p(\tau_1, ..., \tau_p) \prod_{j=1}^p x(n-\tau_j),
\label{eq:volt_general}
\end{equation}
where $x$ is the input signal and $y$ is the output signal. $h_p(\tau_1, ..., \tau_p) \neq 0$, is the $n$-th order discrete kernel of Volterra. The kernel can be considered as a generalization of the impulse response used in linear systems. \\
The parameters $a$ and $b$ represent the memory and they will impact on the filter precision and the performance.\\
The kernel $h_p(\tau_1, ..., \tau_p)$ is symmetric as shown in \cite{nonlineartheory}, this property can be exploited in order to rewrite equation (\ref{eq:volt_general}) in more convenient form, as:

\begin{equation}
\begin{split}
&y(n) = h_0 + \sum_{\tau_1=0}^{M-1} h_1(\tau_1)x(n-\tau_1) + \\
&\sum_{\tau_1=0}^{M-1}\sum_{\tau_2=0}^{M-1} h_2(\tau_1, \tau_2)x(n-\tau_1)x(n-\tau_2) + \\
&\sum_{\tau_1=0}^{M-1} \sum_{\tau_2=0}^{M-1} \sum_{\tau_3=0}^{M-1} h_3(\tau_1, \tau_2, \tau_3) x(n-\tau_1)x(n-\tau_2) x(n-\tau_3)+\\
&...,
 \end{split}
 \label{eq:volt_rewritten}
\end{equation}
where $M$ is the memory length of the model, the limit is given by the interval $[a,b]$ in equation (\ref{eq:volt_general}).\\
In this paper, the series is expanded to third order. The optimization presented in \cite{efficientimplementation}, exploiting the symmetry of the kernels, is used to perform a reduced computation of the Volterra expansion. The optimized version is chosen because it saves memory and performs the computation in a faster way. Equation (\ref{eq:volt_rewritten}) can be rewritten, using the vector representation for all orders, as follows \cite{efficientimplementation}:
\begin{equation}
\begin{split}
y(n) = &\tilde{h_1} (\tau_1, n) x_1^T (n) + \tilde{h_2} (\tau_1, \tau_2, n) x_2^T (n) +\\
&\tilde{h_3} (\tau_1,\tau_2,\tau_3, n) x_3^T(n)
\end{split}
\label{eq:volt_conv}
\end{equation}
where:
\begin{equation}
x_1(n) = [ x(n), x(n-1),  ... , x(n-M+1) ], 
\label{eq:x_first}
\end{equation}
The second order is computed considering only the lower triangular part of the matrix obtained from $x_1(n)^T  x_1(n)$, then
\begin{equation}
\begin{split}
x_2(n) =& [ x^2 (n), x(n)x(n-1), ..., x(n)x(n-M+1),\\
&x^2(n-1), x(n-1)x(n-2),..., x^2(n-M+1) ],
\end{split}
\label{eq:x_second}
\end{equation}
Following the same logic we have
 \begin{equation}
\begin{split} 
 x_3(n) =& [x^3(n), x^2(n)x(n-1),..., x^2(n)x(n-M+1),\\
 &x^3(n-1)x^2(n-1)x(n-2),...,x^3(n-M+1)],
\end{split}
 \end{equation}
$\tilde{h_1}, \tilde{h_2}, \tilde{h_3}$ can be computed similarly\footnote{$\tilde{h_1}(n) = [h_1(0), h_1(1), ..., h_1(M-1)].$}.\\
Using these equations, one can see that the first order vector has $M$ coefficients, the second order vector $M(M+1)/2$ and third order vector $M(M+1)(M+2)/6$ coefficients \cite{computationalcomplexity}.The choice of the value $M$ is extremely important not only for the precision of the model, but also for the performance. \\

\section{Kernel Estimation}
\begin{figure}[H]
\includegraphics[scale=.35]{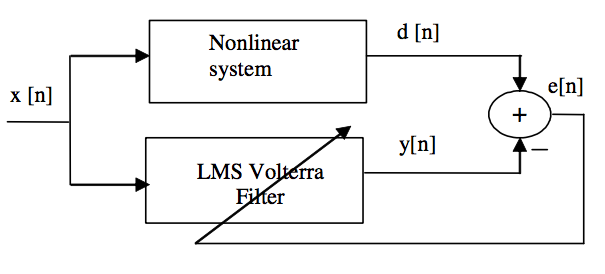}
\caption{Method used to estimate the kernel through the NLMS algorithm. $x$ is the input signal, $d$ is the desired signal, $y$ is the current output signal of the filter and $e$ is the error between $d$ and $y$ which is used to correct the filter. }
\label{fig:lms}
\end{figure}

The Volterra kernels can be computed analytically or through an optimization algorithm. This algorithm is a variation of the well known Least Mean Square (LMS, called Normalized Least Mean Square (NLMS) which has become a standard method for nonlinear system characterization based on Volterra series. In this paper we use optimization algorithm to calculate the kernel coefficients. A schematic representation of the algorithm is shown in Fig. \ref{fig:lms}. 
To estimate the kernels a set of typical input signals is chosen. Each input signal $x$ is sent to loudspeaker and its output is collected with a measurement system. The loudspeaker output represents the desired signal of the model, which we call $d$.\\
The set of input and desired signals represent the knowledge base of the loudspeaker, which will be used to extract the information about the system and build the model. If this set is not representative, the model will not be reliable. To avoid this risk, we have used different signals including monochromatic signals, white noise, chirp and complex signals.\\
The input $x$ is processed with the filter in order to obtain the output $y$, related to current kernels. The correction step is now applied computing the difference between $d$ and $y$, called $e$. $e$ represents the filter error and it is used as weight into correction formula (line 7 of Algorithm \ref{alg:nlms}). So, if the error is large, the correction will be large.\\ 
Algorithm \ref{alg:nlms} explains the entire procedure in a formal way. 
\begin{algorithm}[H]
\begin{algorithmic}[1]
\State{Initialization of $\tilde{h_i}, i=1,.., P$}
\While{$avg>\theta$ }
	\While{$k < l$}
		\State{$y_i(k) = \tilde{h_i} * x_i(k)^T, i=1,..,P$}
		\State{$y(k) = \sum_i y_i(k), i=1,..,P$}
		\State{$e(k) = d(k) - y(k) $}
		\State{$\tilde{h_i} = \tilde{h_i} + \mu_i e(k) x_i(k),  i=1,..,P$}
	\EndWhile
	\State{it = it + 1}
	\State{avg = $\sum_{i=0}^l e(i) / l$}
\EndWhile
\end{algorithmic}
\caption{Algorithm used to estimate Volterra kernels}\label{alg:nlms}
\end{algorithm}
Here $P$ is the filter order, $l$ the length of the signals and $\theta$ the fixed threshold for the precision. $\tilde{h_i}$ is the i-th nonlinear model terms of equation (\ref{eq:volt_conv}) and $\mu_i$ is a learning coefficient of the algorithm defined as:
\begin{equation}
 \mu_i = \frac{\alpha_i}{x_i(n) x_i^T(n) + \phi},
 \label{eq:mu}
\end{equation}
where $\alpha_i$ and $\phi$ are positive constants. $\alpha_i$ should be selected under the constraint $0<\alpha_i < 2$, in order to ensure stability and convergence \cite{adaptivesignalprocessing} and $\phi$ is needed to avoid division by $0$.\\\\
Initializing the Volterra kernels $\tilde{h_i}$ with a random distribution turned out to lead to long convergence time. Instead, using the identity as initialization has led to faster convergence due to the fact that both linear and nonlinear deviations of the system from the identity are small, i.e. for a speaker the input and the out are relatively similar up to a scaling factor. The initialization is made as follows:
\begin{itemize}
\item $h_1$: the first component equal to $1$, the others fixed to $0$
\item $h_2$, $h_3$: all components are fixed to $0$
\end{itemize}

\section{Measurement Technique}

\begin{figure}[H]
\includegraphics[scale=.7]{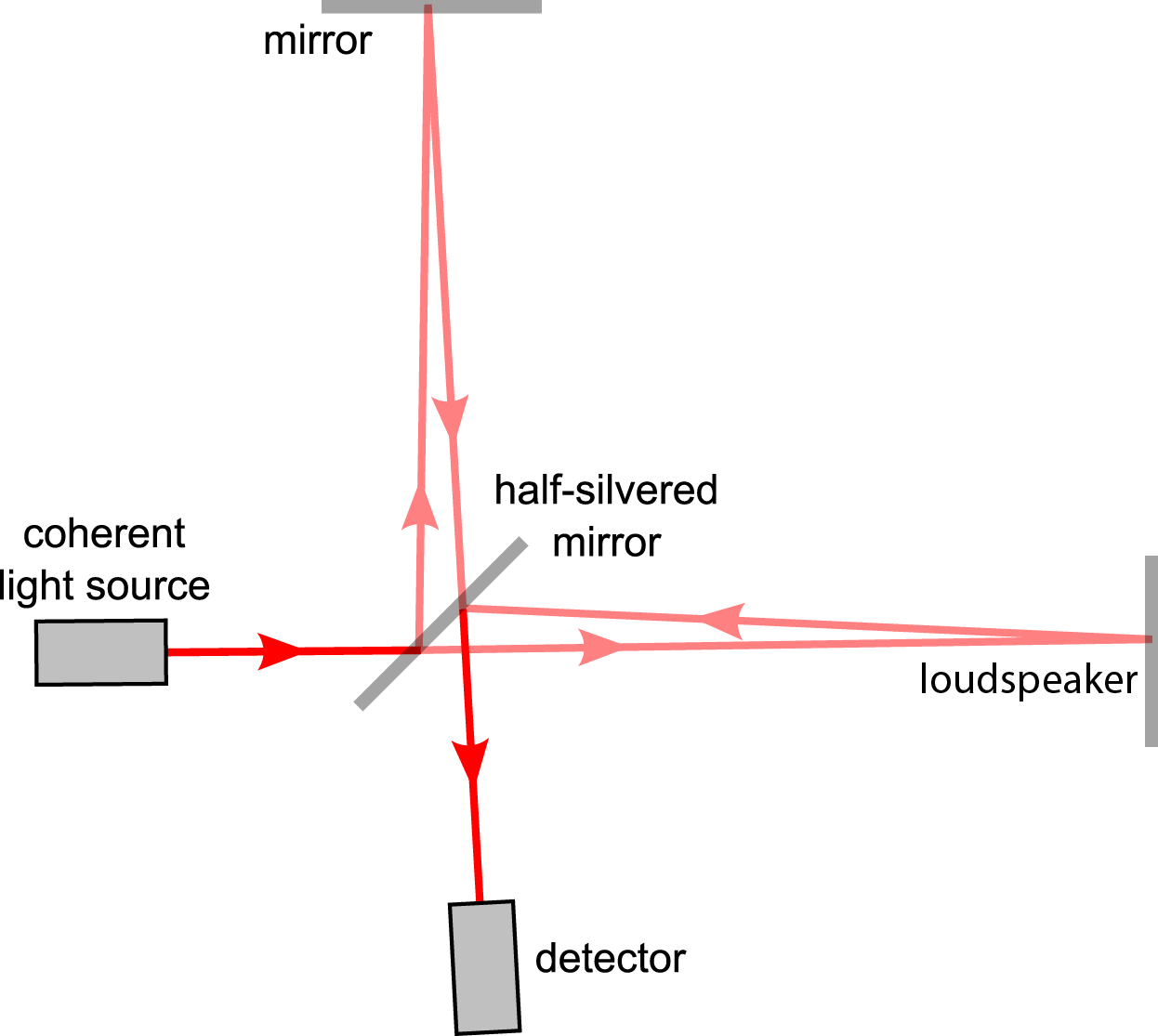}
\centering
\caption{Experimental setup used for the measurement step.}
\label{fig:exp_config}
\end{figure}

The Vibration analysis was performed using a Polytec PSV-400 Scanning Vibrometer.  For single point measurements, the analog output of the laser was connected to OROS OR-38 multianalyzer. For loudspeaker excitation the internal function generator of OR-38 was used. The time signals were processed by OR-38. The experiments were performed in the large experimental hall of the mechanical engineering department of the Zurich University of Applied Sciences. This hall is neither equipped with an anechoic chamber nor acoustically treated. Therefore background noise or reflections from the wall of the building cannot be excluded. The loudspeaker was mounted on top of a heavy vibration table (more than five meters away from any wall) with its cone facing the open part of the room. The laser vibrometer was placed approximately 3\,m from the loudspeaker in order to minimze vibrations of the laser head. A schematic representation can be seen in fig \ref{fig:exp_config}. 
In order to measure the response of the loudspeaker different excitation signals were used: pure sine at different frequencies, sine sweep, periodic chirp and random noise. 

\section{Results}
\label{sec:results}
\begin{figure}[H]
\includegraphics[scale=.35]{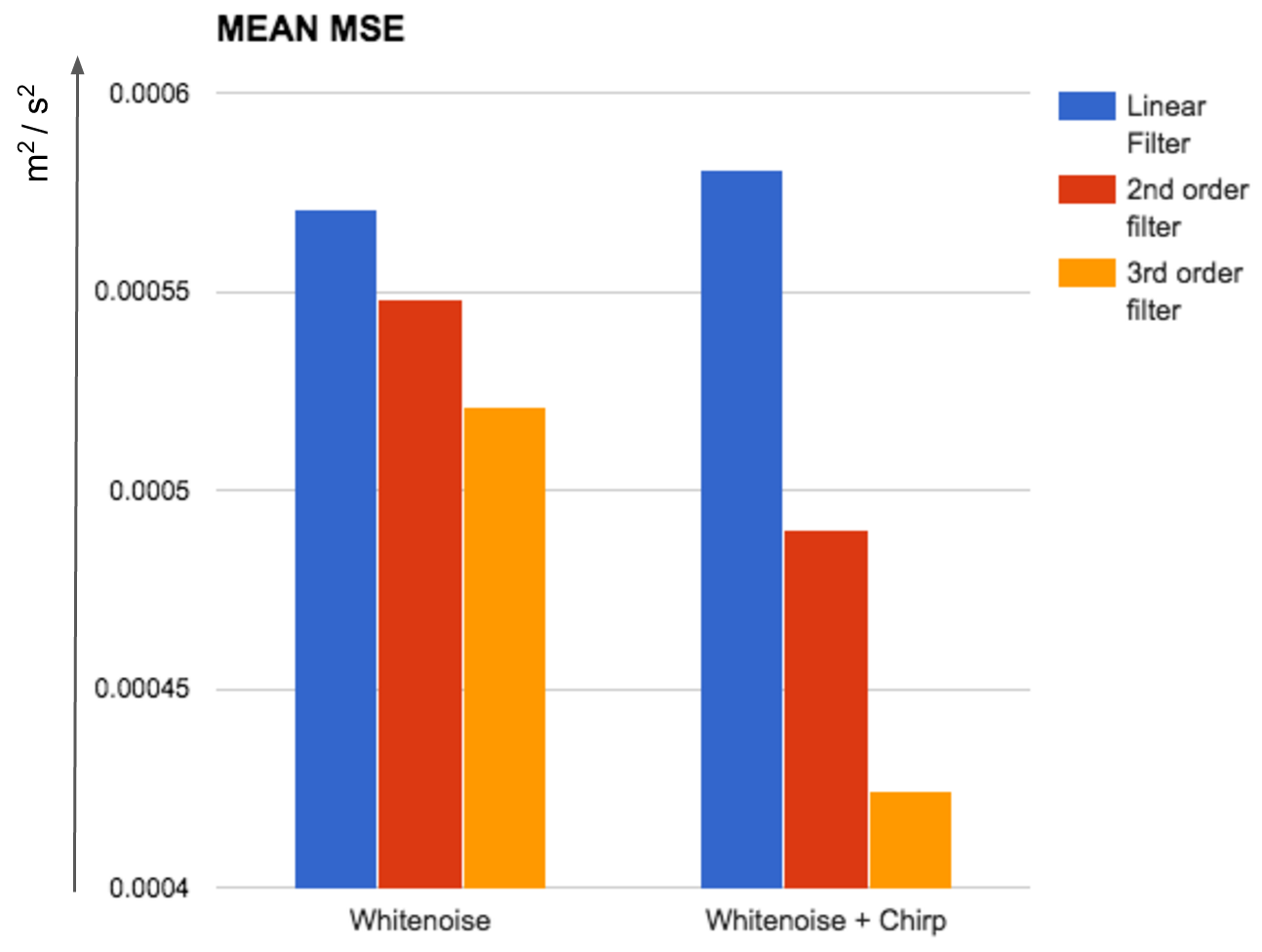}
\caption{Comparison of the average of MSE between the linear, second and third order filter for models trained with white noise and white noise plus chirp respectively.}
\label{fig:comparison_avg}
\end{figure}

\begin{figure}[H]
\includegraphics[scale=.35]{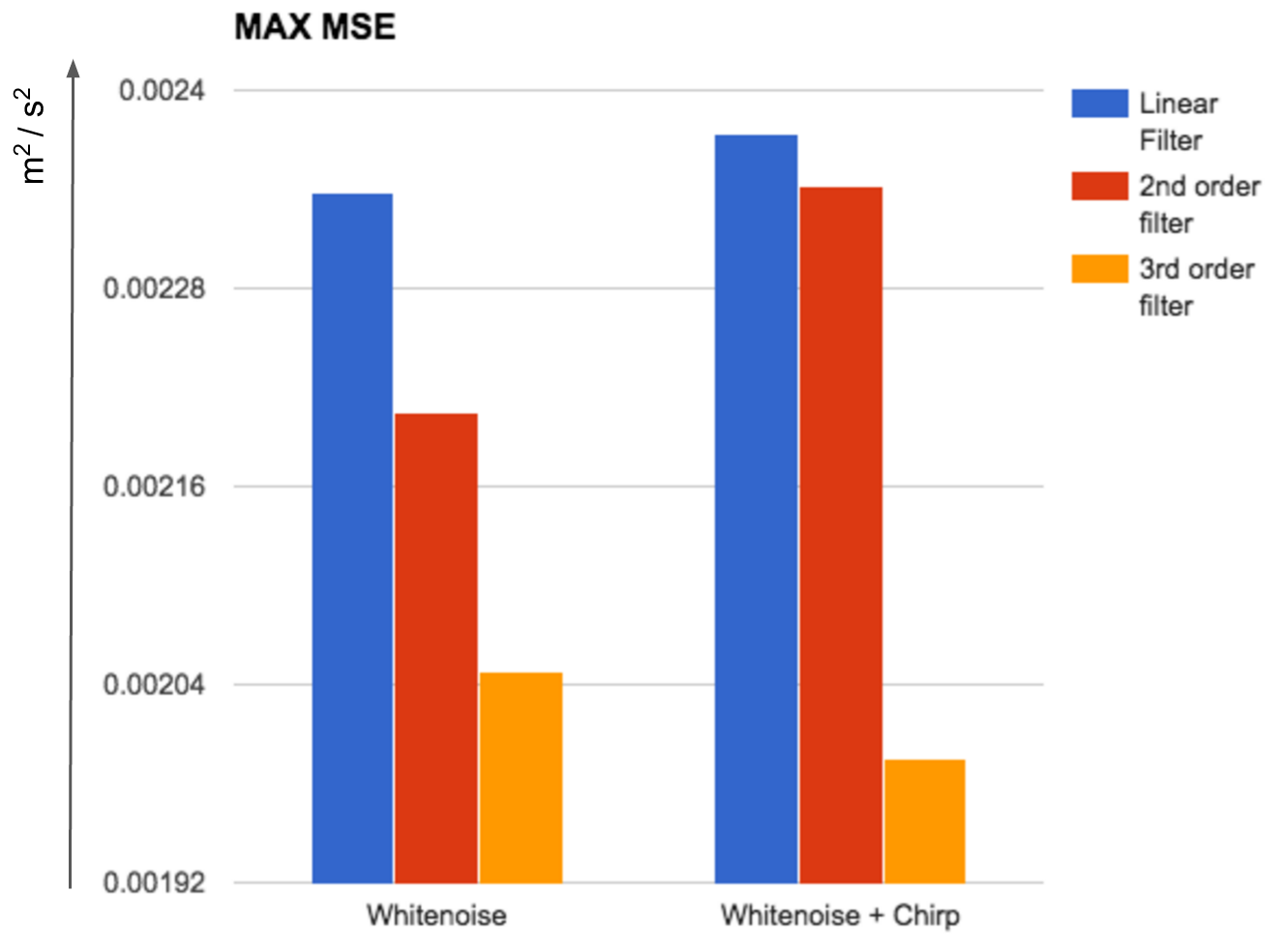}
\caption{Comparison of the maximum of MSE between the linear, second and third order filter for models trained with white noise and white noise plus chirp respectively.}
\label{fig:comparison_max}
\end{figure}

In order to obtain a better knowledge of the Volterra filter capabilities, we tested the trained models with different input signals including chirp, monochromatic and superimposed signals.\\
The benefits of using these signals are as follows: chirp is useful because the frequencies are distributed in the entire spectrum of interest. For this reason, the output signal represents a good approximation of the system behavior in general. Monochromatic signals can show the original frequency (fundamental frequency) and the contribution of non linear terms (harmonics). With the superimposed signals, denoted as \textit{multisine}, we can test intermodulations.\\
The results are summarized in table \ref{tab:output_range} to \ref{tab:comparison_complex}, organized as follows: Table \ref{tab:output_range}, the first column shows the signals used for test and the second column shows the amplitude range for each signal. In Table \ref{tab:comparison_whitenoise} and Table \ref{tab:comparison_complex} a more detailed view is presented. The first column shows the signal used for test and the rest of columns show the MSE of the first, second and third order filter.\\
\begin{table}[h!]
\begin{center}
\begin{footnotesize}
\begin{tabular}{|c|c|}
\hline
\textbf{Signal} & \textbf{Output Range} \\
\hline
CHIRP & [-0.21, 0.27]   \\
\hline
20Hz  & [-0.05, 0.05] \\
\hline
50Hz  & [-0.17,0.17]  \\
\hline
70Hz & [-0.19, 0.19]  \\
\hline
MULTISINE6  & [-0.34, 0.62]   \\
\hline
MULTISINE3  & [-0.45, 0.58] \\
\hline
\end{tabular}
\end{footnotesize}
\end{center}
\caption{Amplitude range of test signals}
\label{tab:output_range}
\end{table}
The signal \textit{multisine6} is composed of superimposed signals where the sub-signals have distance in frequency of $6Hz$, similarly for \textit{multisine3}.\\
For this test we used a model trained only on white noise. The best results have been achieved with memory greater than 60, a point after which the reward for the increasing computation time was decreasing. Results are shown in Table \ref{tab:comparison_whitenoise}.
\begin{table}[h!]
\begin{center}
\begin{footnotesize}
\begin{tabular}{|c|c|c|c|}
\hline
\textbf{Signal} & \textbf{linear} & \textbf{2-th order} & \textbf{3-th order} \\
\hline
CHIRP  & $2.32E-05$ & $2.25E-05$ & $2.10E-05$  \\
\hline
20Hz  &  $7.16E-05$ &	$6.96E-05$ &	$6.62E-05$ \\
\hline
50Hz &  $5.33E-04$ & $0.000535$ & $0.00053928$ \\
\hline
70Hz & $0.0001184$ &	$0.0001142$	& $0.0001132$  \\
\hline
MULTISINE6  & $0.0003404$ & $0.000340$ & $0.0003391$ \\
\hline
MULTISINE3  & $0.0023386$	 & $0.0022054$	& $0.0020482$ \\
\hline
\end{tabular}
\end{footnotesize}
\end{center}
\caption{Comparison between the MSE of linear, second and third order model using white noise as estimation signal. }
\label{tab:comparison_whitenoise}
\end{table}
To improve the model we used a more complex training signal, composed of $60\%$ of white noise and $40\%$ \textit{chirp}. It is our intuition that chirp signal, presenting only one frequency at a time, does train the system in ways that white noise cannot. In this case the diminishing returns point for the memory of the filter was $65$ frames. Results are shown in Table \ref{tab:comparison_complex}.\\
\begin{table}[h!]
\begin{center}
\begin{footnotesize}
\begin{tabular}{|c|c|c|c|}
\hline
\textbf{Signal} & \textbf{linear} & \textbf{2-th order} & \textbf{3-th order} \\
\hline
CHIRP  & $2.38E-05$ & $2.34E-05$ & $2.11E-05$  \\
\hline
20Hz  & $7.59E-05$ & $7.51E-05$ & $7.00E-05$ \\
\hline
50Hz & $0.0005244$ & $0.0005244$ & $0.000479$  \\
\hline
70Hz  & $0.0001333$ & $1.31E-04$  & $1.30E-04$ \\
\hline
MULTISINE6  & $0.000352$  & $0.000334$  & $0.0002713$  \\
\hline
MULTISINE3  & $0.0023742$ & $0.0023423$  & $0.0019952$ \\
\hline
\end{tabular}
\end{footnotesize}
\end{center}
\caption{Comparison between the MSE of linear, second and third order model using a complex signal, composed of white noise and chirp.}
\label{tab:comparison_complex}
\end{table}
Figure \ref{fig:comparison_avg} and \ref{fig:comparison_max} show the average and maximum value of the mean squared error for both experiments. As it can be seen from both figures, higher order models result in a decrease in both maximum and average mean squared error. 
While we expect a more complex model to better model nonlinear distortions, we also observe a trend where white noise plus chirp signal performs better than just white noise as training signal.\\
More specifically, in the case of white nose plus chirp and a third order filter the mean of MSE is $26.95\%$ lower than the MSE obtained form the first order filter. For the same signal and the second order filter this reduction is only $15.55\%$. On the other hand even using a third order filter only achieves a $8.71\%$ reduction when trained on white noise. Considering the max of MSE which is the performance in the worse case, the third order filter achieves an improvement of $15.96\%$ compared to $1.34\%$ of the second order filter. Training a third order filter with only white noise we obtain an improvement of $12.42\%$. 
The results from Table \ref{tab:comparison_complex} show also how the MSE depends on the frequency of the input signal. In particular while the increasing order of the filters pays off for a $70Hz$ sinusoidal, it does not significantly improve the MSE for $50Hz$.

\section{Conclusion}
In this paper, a nonlinear model has been developed in order to simulate the behavior of a loudspeaker. The goal of this study was to asses the simulation improvements for second and third order nonlinear filters compared to linear filters. \\
In order to find a model which improves the accuracy of linear filters preserving good generalization abilities, several signals have been used for testing.
We have shown that white noise can be used to train Volterra kernels with a good accuracy. Furthermore a combination of white noise and chirp has been successfully used to improve the results. We also explored the hyperparameters of the system like memory and learning parameters and shown that the error, expressed as MSE, decreases with higher order of the model. The results also show a dependency on the form and frequency of the signal.\\
Based on this work we are currently performing tests with Volterra filters and neural networks. The results will appear in a forthcoming publication.


\end{document}